\documentclass[useAMS,usenatbib]{mn2e}
\usepackage{epsfig}

\def\be{\begin{equation}}
\def\ee{\end{equation}}
\def\hi{\rm H{\tt I} }
\def\hinsp{\rm H{\tt I}}
\newcommand{\apj}{Ap.\ J.}

\title[Neutral Hydrogen Surveys]{Neutral hydrogen surveys for high
redshift galaxy clusters and proto-clusters} 
\author[R.A.~Battye et~al.]{Richard A. Battye$^1$, Rod D. Davies$^1$
and Jochen Weller$^2$\\$^1$Jodrell Bank Observatory, University of
Manchester, Macclesfield, Cheshire SK11 9DL, U.K.\\$^2$Institute of
Astronomy, Madingley Road, Cambridge CB3 OHA, U.K.} 

\date{Accepted ???, Received ???; in original form \today}
\pagerange{\pageref{firstpage}--\pageref{lastpage}}
\pubyear{2003}

\begin{document}

\maketitle

\label{firstpage}

\begin{abstract}

We discuss the possibility of performing blind surveys to detect
large-scale features of the universe using 21cm emission. Using
instruments with $\sim 5^{\prime}-10^{\prime}$ resolution currently in the
planning stage, it should be possible to detect virialized galaxy
clusters at intermediate redshifts using the combined
emission from their constituent galaxies, as well as less overdense
structures, such as proto-clusters and the `cosmic web', at higher
redshifts.  Using semi-analytic methods we compute the number
of virialized objects and those at turnaround which might be detected
by such surveys.  We find a surprisingly large number of objects might
be detected even using small ($\sim 5\%$) bandwidths and elaborate on
some issues pertinent to optimising the design of the instrument and
the survey strategy. The main uncertainty is the fraction of neutral
gas relative to the total dark matter within the object. We discuss
this issue in the context of the observations which are currently available.

\end{abstract}


\section{Introduction}

Emission and absorption of electromagnetic radiation due to
transitions between the hyperfine states of neutral Hydrogen
(\hinsp)~\citep{field} has had a significant impact on our understanding of
our galaxy and our immediate cosmic neighbourhood.  At
present, however, the highest redshift detection of 21cm emission is
at $z\approx 0.18$~\citep{Zwann:2001a} and only very shallow surveys
$(z<0.04)$ of the whole sky have been
performed~\citep{Zwann:1997a,Kilborn:1999a,Ryan-Weber:2001a,Lang:2003a}.
In this {\em article} we would like to point out that it might soon be
possible to perform blind unbiased searches for large objects, both
virialized and collapsing, using 21cm emission as 
their tracer.

It has long been speculated that it might be possible to detect the
onset of the epoch of reionization using redshifted 21cm
emission~\citep{rees1,rees2} and indeed, spurred on by the claim that
the reionization epoch might have been at much higher redshifts
($z\approx 17$) than previously thought~\citep{spergel}, there has
been much recent work on this
subject~\citep{furl1,zald1,gnedin}. Moreover, it has also been pointed
out that the evolution of the spin-temperature, $T_{\rm S}$, prior to
reionization might be such that the large-scale structure can be
observed in absorption against the cosmic microwave background (CMB)
at very low frequencies ($\sim 30-50\,{\rm MHz}$) enabling accurate
determination of a variety of important cosmological
parameters~\citep{zald2}. Our motivation is somewhat different: to map
the large-scale structure  (LSS) once it has developed sufficiently
for the spin temperature to be given by the kinetic temperature of the
gas $\sim 1000{\rm K}$ and $T_{\rm S}\gg T_{\rm CMB}$, that is, using
21cm emission. 

For various proposed instruments, our aim is to make conservative
estimates of the number of virialized clusters that might be found in
the range $z<1$, and also collapsing objects with lower overdensities
at $z>1$. These higher redshift objects are the proto-clusters which
would have formed virialized clusters by the present day and in some
sense characterize the so called `cosmic web' of LSS observed in
cosmological N-body simulations. They are likely to be \hi rich, but
naively, virialized objects would not necessarily be the most obvious
places to look for neutral gas since the process of virialization and 
the creation of the intracluster medium involves violent gravitational
processes which are likely to strip neutral gas from the constituent
galaxies by tidal interactions and ram pressure~\citep{ram}. However,
by virtue of their large overdensity such objects should still contain a
substantial neutral component, probably larger at high redshifts than
locally, albeit locked up in galaxies. The fiducial detection
threshold that we will discuss in this paper will be $\approx
10^{11}M_{\odot}$ of \hinsp, which we will show could be possible on
large patches of the sky; it is worth pointing out that this would
only require a cluster to contain 20 gas-rich spirals with \hi masses
$\sim 5\times 10^{9}M_{\odot}$. This is likely to be the case for many
clusters. 

Most importantly, in the context of this discussion, all the objects
we will consider, both virialized and collapsing, will have angular
diameters $\sim 
5^{\prime}-10^{\prime}$ which will match the resolution of the telescopes and
arrays with $\sim 100$m baselines currently being planned to detect
redshifted 21cm emission. This will make them ideal survey targets
since they would fill the beam, in contrast to ordinary galaxies at
the same redshift whose signal will be substantially diluted and will
most likely be below the confusion limit. 

Blind surveys for galaxies and galaxy clusters have been performed in
the optical waveband for many years. The most recent galaxy redshift
surveys (2dFGRS and SDSS) have yielded significant constraints on
cosmological parameters via measurements of the power spectrum and
redshift space distortions, as well as a wealth of understanding of
the properties of the individual galaxies~\citep{Peacock:2003a,teg1}. If deep 
redshift surveys using \hi could be performed they would open up a new window
on the universe which could have some technical advantages over optical
approaches. In particular, using the flexible spectral resolution of
radio receivers, objects can be selected using spectroscopic methods
rather than, for example, photometrically from optical plates,  making
them flux limited over the specific redshift range probed. The
structure of the selection function should, therefore,  be 
simple to define. It should also be possible to deduce the redshift and 
accurate line-widths for individual objects allowing one to probe the
underlying mass distribution directly and establish the precise
selection 
criteria a posteriori. This could overcome some of the issues of bias
inherent in optical surveys. Moreover, potential confusion with other
spectral lines is unlikely in the radio waveband since there a very
few strong lines. 

A standard line of argument often put forward is that, the optical
surveys are sensitive to the sum of all the starlight from the
galaxies and could be significantly biased by the non-linearities
involved in the star-formation process, while the neutral component is
of primordial origin and hence is unbiased. This is unlikely to be
completely true since the neutral component is also involved in these
processes. However, observing the \hi component is  more than just
another view of the same object. The \hi halos of nearby galaxies
extend out very much further than their optical counter parts,
suggesting that the dynamics of the \hi contains extra information. 

As with galaxies, the study of clusters has focused on the central
region where diffuse hot gas is situated. This emits very strongly in
the X-ray waveband and can also lead to fluctuations in the CMB via
the Sunyaev-Zeldovich (SZ) effect. However, typically the core radius
of the cluster is only about 1/10th of linear size of the object and
1/1000th of the volume. It is likely that the study of these objects
using \hi as the tracer will lead to new insights in an analogous way
to galaxies. In particular many of the \hi rich spiral galaxies are
likely to be situated away from the core and there is also a
possibility of a component of diffuse \hi similar to that seen around
individual galaxies. 

At  present radio telescopes which are sufficiently sensitive
to map the sky quickly enough to be competitive with optical surveys
do not exist. However, the next few years could see the start of
development of instruments which are many times more powerful in terms
of survey speed than those presently available. The Square Kilometre
Array (SKA) (see {\tt www.skatelescope.org} for details) is the
ultimate goal of much technology development currently
underway. Although its precise design is still evolving, its
projected specification involves a collecting area of $\sim 10^{6}{\rm
m}^2$, sub-arcsecond resolution, a large instantaneous field of view
and significant bandwidth in the region where redshifted \hi will be
detected. Such an instrument will have the ability to perform
significant galaxy redshift surveys tracing \hi over a wide range of
redshifts. 

Proto-types are being planned which would be around 1/100th of the
full SKA in area. Searches for galaxies at low redshift, extending
those currently being performed, would be an obvious possibility for
such an instrument. Our aim, here, is to discuss issues related to the
design of such an instrument in the context of a blind survey for
large objects at intermediate and high redshifts such as virialized
galaxy clusters and collapsing proto-clusters which are, as we have
already discussed the ideal targets for these surveys in terms of
their size; our conclusion being that the planned instruments will be
sufficiently sensitive to perform significant surveys. 

First, we will discuss different proposed concepts for such
instruments and perform a simple sensitivity calculation for the
limiting \hi mass ($M_{\hi}^{\rm lim}$) of a survey assuming a fixed
integration time\footnote{We note that all integration times discussed
in this paper are `on-source' integration times and do not take into
account any practical difficulties required to make the
observations. These could vary significantly for different
experimental configurations.}.  We will then discuss the possibility
of detecting virialized clusters. In order to make the link with
theoretical predictions of the number of dark matter halos, we need to
make some assumption as to  the \hi content of individual dark matter
halos. This is a very complicated issue and since we are, at this
stage, only trying to make rough estimate of the number of clusters
one might find,  we will assume that there exists a universal ratio
$f_{\rm H\tt I}(M,z)=M_{\hi}/M$ which relates the dark matter mass of
a halo, $M$,  to its neutral gas component at a given redshift. This
should be a good approximation in the  mass range relevant to
clusters, but is unlikely to be so of smaller masses. We will make
various arguments in order to estimate this fraction. At each stage we
will attempt to make conservative assumptions and therefore our
estimate of the number of objects found in a given survey should be a
lower bound. Under these assumptions we can then deduce the number of
objects that a survey might find using the results of N-body
simulations. We then discuss some issues related to the design of the
instrument, the survey strategy and confusion due to \hi in the
field. Finally, we will adapt our calculations to compute the number
of objects at turnaround and discuss issues relating to their
detection   

Throughout this paper, we will assume a cosmological model which is
flat, with the densities of matter and the cosmological constant
relative to critical given by $\Omega_{\rm m}=0.3$ and
$\Omega_{\Lambda}=0.7$. The Hubble constant used is $72\,{\rm km}\,{\rm
sec}^{-1}{\rm Mpc}^{-1}$ and the power spectrum normalization is
assumed to be $\sigma_8=0.9$. These values are compatible with a range
of observations~\citep{spergel,contaldi:2003a}.

\section{Proposed instruments}

A number of different concepts are under discussion which could make
observations in the frequency range  $400\,{\rm MHz}<f_{\rm
obs}<1200\,{\rm MHz}$. The proposed instruments will, in fact, be
sensitive to a much wider range of frequencies (up to $1700\,{\rm
MHz}$), but this is the most sensible range to discuss the detection
of large objects since they will be close to, or completely,
unresolved. The design features which are of interest to us here  are
the overall collecting area, $A$, the aperture efficiency, $\eta$, the
instantaneous bandwidth, $\Delta f_{\rm inst}$, the instantaneous
field-of-view (FOV), $\Omega_{\rm inst}$, the longest baseline, $d$,
and the system noise temperature of the receivers, $T_{\rm sys}^{\rm
rec}$. The overall system temperature, $T_{\rm sys}$,  at frequency
$f$ can be conservatively estimated by taking into account the CMB and
galactic background  
\be
T_{\rm sys}\approx T_{\rm sys}^{\rm rec}+2.7{\rm K}+10{\rm
K}\left({f\over 408{\rm MHz}}\right)^{-3}\,, 
\ee
in a cold part of the sky.

We will focus on two different concepts which differ in the way they
form the instantaneous field of view. Firstly, we will consider a
conventional interferometer similar to the Allen Telescope Array
(ATA), which comprises $n_d$ dishes of diameter $D$. In this case
$\Omega_{\rm inst}\propto (\lambda/D)^2$ and $A=\pi n_{\rm d} D^2/4$
where $\lambda\propto 1/f_{\rm obs}$. The current design of the ATA
has $n_{\rm d}\approx 350$, $D=6.1{\rm m}$ and it is likely that
$\eta\approx 0.6$, $d\approx 400{\rm m}$ and $\Delta f_{\rm inst}/
f_{\rm obs}\approx 0.05$. The FOV is set by the  size of the
individual telescopes and the synthesized beam size is set by the
longest baseline. Such a configuration would have a synthesized beam
with $\theta_{\rm FWHM}=3.2^{\prime}$ at $f_{\rm obs}=1000{\rm MHz}$
and a corresponding instantaneous FOV of $13.4\,{\rm deg}^2$. Note
that throughout we assume a circular aperture, that is, $\theta_{\rm
FWHM}=1.22\lambda/d$, which should be a reasonable approximation.  

The other competing concept which we will consider is that of a phased
array which could potentially have a much larger field of view; the
idea being to use very low cost antennae, arranged in $n_{\rm p}$
patches of size $P$, and form $n_{\rm b}$ individual beams using
off-the-shelf computer hardware. Hence, $\Omega_{\rm inst}\propto
n_{\rm b}(\lambda/d)^2\le (\lambda/P)^2$ and $A=\pi n_{\rm p}P^2/4$
assuming that the patches are circular, from which we can deduce that
$d\ge \sqrt{n_{\rm b}}P$. Such a concept could have more than one
instantaneous FOV, possibly allowing for more continuous usage for
$\hi$ studies. Three possible scenarios for such an instrument are
presented in table~\ref{tab:tab1}. These range from conservative,
setup I, to speculative, setup  III. We should note that for $P\approx
4{\rm m}$, none of the above configurations come close to saturating
the inequality on $P$. The value of $n_{\rm b}$ should  be taken as
being a guide figure since extra computing power could increase this
close to the maximum FOV set by $P$. 

\begin{table}
\begin{center}
\begin{tabular}{|l|c|c|c|}
\hline
   & I & II & III\\
\hline
$A/m^2$  & 5000 & 10000 & 15000 \\
$\Delta f_{\rm inst}/f_{\rm obs}$ & 0.02 & 0.05 & 0.10 \\
$d/m$ & 150 & 250 & 350 \\
$n_{\rm b}$ & 1000 & 2000 & 4000 \\
$T_{\rm sys}^{\rm rec}/K$ & 40 & 30 & 20 \\
$\theta_{\rm FWHM}$ & $8.4^{\prime}$ & $5.0^{\prime}$ & $3.6^{\prime}$ \\
$\Omega_{\rm inst}/{\rm deg}^2$ & 22.2 & 15.9 &  16.3\\
${\cal M}$ & $\approx 5$ & $\approx 60$ & $\approx 550$ \\
\hline
\end{tabular}
\end{center}
\caption{Three possible sets of design parameters for a phased
array. Included also are $\theta_{\rm FWHM}$ and $\Omega_{\rm inst}$
at $f_{\rm obs}=1000\,{\rm MHz}$, and the figure of merit ${\cal M}$
relative to configuration described in the text. In each case we will
assume that $\eta\approx 0.8$. This value, which is slightly larger
than that conventionally assumed for parabolic dishes, should be
possible for these arrays since they are much closer to the ground.} 
\label{tab:tab1}
\end{table}

The standard formula~\citep{Roberts:1975a} relating the \hi mass,
$M_{\hi}$, and the observed  flux density, $S$, integrated over a
velocity width, $\Delta {\rm v}$, in an FRW universe is given by  
\be
{M_{\rm HI}\over M_{\odot}}={2.35\times 10^{5}\over 1+z}\left({d_{\rm
L}\over {\rm Mpc}}\right )^2\left({S\Delta {\rm v}\over {\rm Jy}\,{\rm
km}\,{\rm sec}^{-1}}\right)\,, 
\label{msrel}
\ee
where $d_{\rm L}(z)$ is the luminosity distance to redshift $z$. For a
fiducial velocity width of $\Delta {\rm v}\approx 800\,{\rm km}\,{\rm
sec}^{-1}$ assumed at this stage to be independent of the mass, an
object containing $10^{11}M_{\odot}$ of \hi will have $S_{800}\approx
4,\,14,\,42,\,150,\,840\,{\mu\rm Jy}$ at $f_{\rm
obs}=400,\,600,\,800,\,1000,\,1200\,{\rm MHz}$ corresponding to
$z=2.55,\,1.37,\,0.78,\,0.42,\,0.18$ respectively. 

\begin{figure}
\epsfig{file=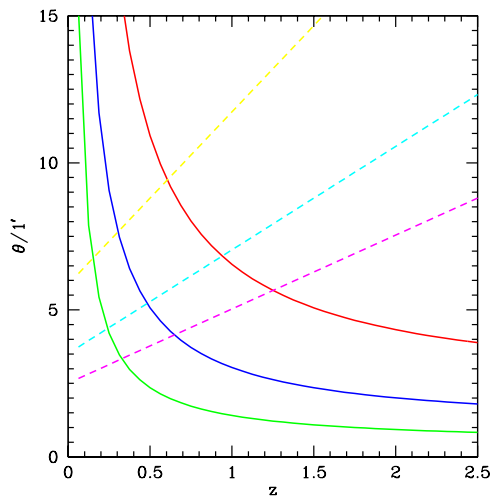,height=7cm}
\caption{The angular diameter size (twice the virial radius) of
virialized objects with $M_{\rm vir}=10^{15}M_{\odot}$,
$10^{14}M_{\odot}$ and $10^{13}M_{\odot}$ (solid lines, top to bottom
respectively). Included also is the estimated beam size for the three
different phased array configurations (dashed lines, I top, II middle,
III bottom). Only the closest and largest objects will be resolved by
the instruments discussed in this paper.} 
\label{fig:theta}
\end{figure}
Assuming that the clusters are unresolved, which will be a good
approximation at least for the phased array configurations at
intermediate and high redshift (see fig.~\ref{fig:theta} for
virialized objects), then the rate at which one can survey the sky
over the instantaneous bandwidth is quantified by  
\be 
{\cal R}_{800}={2T_{\rm sys}\over \eta A \sqrt{\Delta f_{\rm
obj}\Omega_{\rm inst}}}\,, 
\label{rdef}
\ee
where $\Delta f_{\rm obj}$ is the frequency interval corresponding to
$\Delta {\rm v}$. The noise, $S_{\sigma}$, attainable for a survey
with angular coverage, $\Delta\Omega$, in integration time, $t$, is
given by $S_{\sigma}={\cal R}_{800}\sqrt{\Delta\Omega/t}$. This line
of argument should also yield important information even when the
object is resolved. Projected values for ${\cal R}_{800}$ are
presented in table~\ref{tab:tab2} for the ATA and the different phased
array setups using $f_{\rm obs}=400-1200\,{\rm MHz}$ along with the
integration time required to achieve a 5$\sigma$ detection
$(S_{800}>5S_\sigma)$ of $M_{\hi}^{\rm lim}\approx 10^{11}M_{\odot}$
of $\hi$ in $\Delta {\rm v}\approx 800\,{\rm km}\,{\rm sec}^{-1}$ on
$100\,{\rm deg}^2$. Note that in reality the angular coverage of a
given survey is likely to be an integer multiple of the instantaneous
FOV. However, this way of presenting the sensitivity makes it possible
to compare experiments with different FOVs in a coherent way. 
\begin{table}
\begin{center}
\begin{tabular}{|l|c|c|c|c|c|}
\hline
 & $f_{\rm obs}/{\rm MHz}$ & ATA  & I & II & III\\
\hline

& 400 & 1.80 & 2.98 & 1.43 & 0.73 \\
& 500 & 1.74 & 3.01 & 1.41 & 0.69 \\
 & 600 & 1.78 & 3.14 & 1.45 & 0.69  \\
& 700 & 1.85 & 3.31 & 1.51 & 0.71 \\
${\cal R}_{800}$ & 800 & 1.93 & 3.48 & 1.59 & 0.74 \\
& 900 & 2.02 & 3.66 & 1.67 & 0.78 \\
 & 1000 & 2.11 & 3.84 & 1.74 & 0.81 \\
& 1100 & 2.20 & 4.01 & 1.82 & 0.84 \\
& 1200 & 2.29 & 4.18 & 1.89 & 0.88\\
\hline
& 400 & 4480 & 12000 & 2830 & 730 \\
& 500 & 1335 & 4000 & 875 & 200 \\
& 600 & 460 & 1445 & 300  & 70 \\
& 700 & 165 & 535 & 110 & 25\\
$t_{800}$ & 800 & 60& 190  & 40 & 8.8  \\
& 900 & 20 & 64 & 13 & 2.8\\
  & 1000 & 5.7& 19 & 3.9  & 0.9 \\
& 1100 & 1.4 & 4.5& 0.9 & 0.2\\
& 1200 & 0.2 & 0.7 & 0.15 & 0.03\\
\hline
\end{tabular}
\caption{Approximate survey sensitivity ${\cal R}_{800}$ for $f_{\rm
obs}=400-1200\,{\rm MHz}$ in units of ${\rm mJy}\,{\rm
sec}^{1/2}\,{\rm deg}^{-1}$ for the ATA and the phase array setups I,
II and III. Also presented is the integration time $t_{800}$ in days
required to achieve a $5\sigma$ detection threshold of $M_{\hi}^{\rm
lim}=10^{11}M_{\odot}$ on $100\,{\rm deg}^2$ and $\Delta {\rm
v}=800\,{\rm km}\,{\rm sec}^{-1}$.} 
\end{center}
\label{tab:tab2}
\end{table}

\section{Detecting galaxy clusters}

\subsection{\hi content of galaxy clusters}

As is almost always the case, the observed quantity $M_{\hi}$ is not
theoretically well understood. It is possible to make accurate
predictions as to the clustering of dark matter, but the evolution of
the gas is much more difficult to predict. Therefore, we will
introduce the empirical quantity $f_{\hi}(M,z)$ which we will define
to be fraction of the \hi in a dark matter halo of mass $M$ at
redshift $z$. We anticipate that this quantity is likely have a
substantive 
scatter $\sigma_{\hi}^2=\langle f_{\hi}^2\rangle-f_{\hi}^2$, but 
we shall assume this is zero since  estimating it
would be futile due to the small number of the
observations available at present, and any scatter is likely to
increase the number of objects that are found in a given survey
assuming that the mean value is correct. This 
is because there are typically many more objects just below the mass limit than
above it. We will first  present a simple order of magnitude estimate
for $f_{\hi}$ and then go on to discuss the current status of
observations.  
\begin{figure}
\epsfig{file=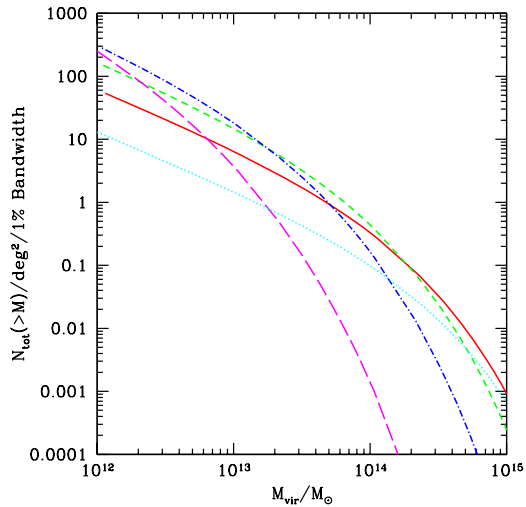,height=7cm}
\caption{The number of dark matter halos with mass greater than $M_{rm
vir}$ per ${\rm deg}^2$ per $1\%$ bandwidth. The dotted line is for
$f_{\rm obs}=1200\,{\rm MHz}$, the solid line is for $1000\,{\rm
MHz}$, the short dashed line for $800\,{\rm MHz}$, the dot-dashed line
for $600\,{\rm MHz}$ and the long dashed line for $400\,{\rm
MHz}$. Note that there is between 1 and 0.1 halo with $M_{rm
vir}\approx 10^{14}M_{\odot}$ for $f_{\rm obs}\ge 600\,{\rm MHz}$.} 
\label{fig:hi}
\end{figure}

One can estimate the fraction of \hi in clusters, $f_{\hi}^{\rm
clust}$, from that in a typical galaxy, by taking into account
different mass to luminosity ratios, $M/L$, of typical  clusters and
galaxies, and the fact that only spiral galaxies are $\hi$ rich. If
$f_{\rm spiral}$ is the fraction of spiral galaxies in a typical
cluster then  
\be 
f_{\rm HI}^{\rm clust}\sim f_{\rm spiral}{(M/L)^{\rm gal}\over
(M/L)^{\rm clust}}f_{\rm HI}^{\rm gal}\,. 
\label{hiest}
\ee
By considering the average properties of
galaxies~\citep{Roberts:1994a}, the value of $f_{\rm HI}^{\rm gal}$ is
observed to be $\sim 0.06$ for a typical spiral galaxy at $z\approx
0$. The value $M/L$ of such object is likely to be $\sim 8$, whereas
that for a typical cluster is $\sim 240$. It is believed that the
value of $f_{\rm spiral}\sim 0.5$, making  $f_{\rm HI}^{\rm clust}\sim
10^{-3}$. 

Another approach is to compute the value of $f_{\rm HI}$ for a cluster
by summing up the measured values of $M_{\hi}$ for the constituent
galaxies and computing their line of sight velocity dispersion,
$\sigma$, and hence their virial mass $M_{\rm vir}=3\pi\sigma^2R_{\rm
H}/(2G)$, where $R_{\rm H}$ is the mean harmonic radius and $G$ is the
gravitational constant. We should note that this is likely to
under-estimate $f_{\rm HI}$ since it ignores the many gas-rich dwarf
galaxies which are below  the detection threshold of any
survey. Moreover, it also ignores, as does the order of magnitude
estimate above, the possibility that there exists a diffuse component
of \hi such as the tidal plumes of \hi observed in galaxy
groups~\citep{Appleton:1981a}. 

The \hi content of the Virgo galaxy cluster is the best
 studied~\citep{Hutchmeier:1986a}, but it is well known that this object is not
 completely relaxed; it being comprised of three interacting
 sub-clusters, and so the computed value of $f_{\hi}$ is likely to be
 an unreliable statistical indicator of the global value.
 Studies of the \hi content of other clusters are much less detailed.
 The Abell cluster A3128 has been surveyed for
 \hi~\citep{Chengular:2001}. By co-adding the \hi at the position of
 galaxies brighter than an r-magnitude of 16.2, a total \hi mass of
 $1.3\times 10^{11}M_{\odot}$ was detected. $M_{\rm vir}$ for this
 cluster is $\approx 1.5\times 10^{14}M_{\odot}$ estimated from the
 velocity dispersion and the angular distribution of the galaxies at
 an assumed distance of 240Mpc. For the detected galaxies
 $M_{\hi}/M_{\rm vir}\approx 0.9\times 10^{-3}$, compatible with our
 earlier order of magnitude estimate. A recent sensitive multi-beam
 study of Centaurus A group has been made \citep{banks:1999a} and  this
 survey was able to detect \hi in galaxies over a wide range of
 luminosities and, therefore, includes at least part of the dwarf
 galaxy contribution. For this group $M_{\hi}/M_{\rm vir}\approx
 1.1\times 10^{-3}$. 
 \begin{figure}
\epsfig{file=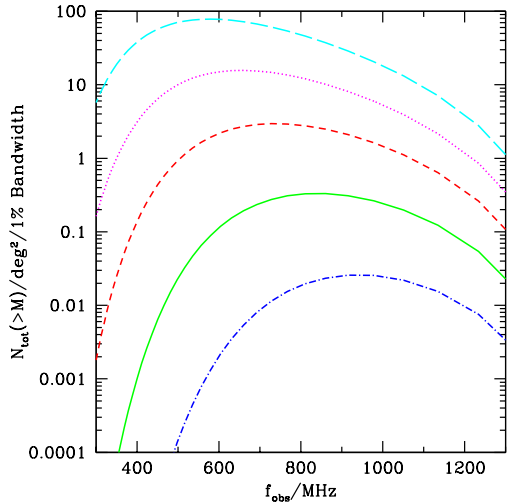,height=7cm}
\caption{The same quantity as plotted in fig.~\ref{fig:hi} but this
time against $f_{\rm obs}$ for $M_{\rm vir}=3\times 10^{12}M_{\odot}$
(long-dashed line),  $10^{13}M_{\odot}$ (dotted line),  $3\times
10^{13}M_{\odot}$ (short-dashed line), $10^{14}M_{\odot}$ (solid line)
and $3\times 10^{14}M_{\odot}$ (dot-dashed line).} 
\label{fig:dn}
\end{figure}

Many different arguments suggest that $f_{\hi}$ should increase with
redshift. \hi is the basic fuel for star formation and it is believed
that for an individual object $\dot M_{\star}\propto M_{\hi}$. Since the
global star formation rate is known to grow approximately linearly
with redshift to a maximum about five times the local value at
$z\approx 2-3$, it seems likely that $f_{\hi}^{\rm gal}(M,z\sim
1)\approx 5\times f_{\hi}^{\rm gal}(M,z\sim 0)$; an effect which
may be even stronger within clusters. A related, but logically
distinct fact is the observation that high redshift clusters contain a
larger fraction of blue galaxies when compared to
local ones~\citep{butcher:1984a} and also that they contain a larger fraction
of spirals and a reduced number of S0s~\citep{dressler:1997a}. Both these
observed facts suggest that $f_{\rm spiral}$ also increases with
$z$. Hence, using the estimate (\ref{hiest}) it seems sensible to
deduce that using the locally observed $f_{\hi}^{\rm clust}$ is likely
to be an under-estimate by a factor of a few. 

Various estimates of $\Omega_{\hi}$ appear in the literature and these
can be used to gain some further insight into the issues under
consideration here. By making observations of damped Lyman-$\alpha$
systems in an average redshift range of $\langle z\rangle\approx 2.4$
the CORALS project~\citep{Elison:2001a} estimate that
$\Omega_{\hi}\approx 2.6\times 10^{-3}$. Using the fiducial value of
$\Omega_{\rm m}$, one can deduce a value of $f_{\hi}\approx 8\times
10^{-3}$ for the field at this epoch corresponding to  a density of
$\rho_{\hi}\approx 3.6 \times 10^{8}M_{\odot}{\rm Mpc}^{-3}$. Using
the recently published data from the HIPASS survey it has been deduced
~\citep{Zwann:2003c} that for $z<0.04$ the value of $\Omega_{\hi}$ is
much lower,  $\Omega_{\hi}\approx 3.8\times 10^{-4}$ from which one
can deduce that $f_{\hi}\approx 1.2\times 10^{-3}$ and
$\rho_{\hi}\approx 5.3\times 10^{7}M_{\odot}{\rm
Mpc}^{-3}$. Remarkably this is approximately a factor 6-7 smaller than
the estimated value at $z=2.4$ and is clearly compatible with our
earlier remarks concerning the global star formation rate. Moreover,
the estimated value at $z\approx 0$ is very close to the value we have
discussed in the context of clusters. This would tend to suggest that
the value of $f_{\hi}$ is only marginally lower in clusters  (say
$20-30\%$) than in the field. Since the clusters are a substantial
$(\sim 100)$ overdensity in the dark matter they should be clearly
observable in \hi with an appropriate sized telescope. 

\subsection{Number of virialized dark matter halos}

Since we are only trying to compute a lower bound on the number of
objects found in a given survey, it is reasonable to assume a value of
$f_{\rm HI}\approx 10^{-3}$ constant over the range of masses we are
considering and independent of redshift.  If  $M_{\hi}^{\rm
lim}\approx 10^{11}M_{\odot}$ as discussed earlier for the various
planned surveys,  then $M_{\rm lim}\approx 10^{14}M_{\odot}$ for the
dark matter and hence by computing the number of objects greater than
this limit will give an estimate of the number of objects likely to be
found in a survey. 

Now, let us compute the number of objects with mass greater than some
limiting mass $M_{\rm lim}$ between $z-\Delta z/2$ and $z+\Delta z/2$
in a solid angle $\Delta\Omega$,  
\be
N(M>M_{\lim})=\Delta z \Delta \Omega {dV\over
dz\,d\Omega}\int^{\infty}_{M_{\rm lim}} {dn\over dM}dM\,, 
\label{ndef}
\ee
where 
\be
\Delta z=(1+z)d{\rm v}=(1+z)\Delta f_{\rm inst}/f_{\rm obs}\,,
\label{dz}
\ee
is assumed to be small, $dV/(dz\,d\Omega)$ is the comoving volume
element and $dn/dM$ is the comoving number density of objects.

We will use an expression for the comoving number density which is
derived from numerical simulations carried out by the VIRGO
consortium~\citep{Evrard:2002a}  
\be 
{dn\over dM}(z,M) = -0.22 {\rho_{\rm m}(t_0) \over M\sigma_M}
{d\sigma_{\rm M}\over dM} \exp\{-|A(z,m)|^{3.86}\} \,, 
\label{dndM}
\ee
where $A(z,M)=0.73-\log[D(z)\sigma_M]$, $\rho_{\rm m}(t_0)$ is the
matter density at the present day, $\sigma_M$ is the overdensity in a
virialized region of mass $M$ and $D(z)$ is the normalized growth
factor. Here, the mass is defined to be $M_{200}$, that inside a
region with an approximately spherical overdensity $\Delta_{\rm
c}=200$. This can be related to the virial mass, $M_{\rm vir}$ if we
assume an NFW function for the cluster profile~\citep{NFW}. In all the
subsequent discussion we have made this correction using a
concentration parameter of $c=5$, in which case that from $M_{200}$ to
$M_{\rm vir}$ is typically very small. 

In Fig.~\ref{fig:hi}, we plot $N(M>M_{\rm lim})$ per ${\rm deg^2}$ per
$1\%$ bandwidth against $M_{\rm vir}$ for five different redshifts
(z=2.55, 1.37, 0.78, 0.42, 0.18) corresponding to frequencies $f_{\rm
obs}=400\,{\rm MHz}$, $600\,{\rm MHz}$,  $800\,{\rm MHz}$, $1000\,{\rm
MHz}$ and $1200\,{\rm MHz}$ respectively.  We see that there are $\sim
0.5$ dark matter halos with $M_{\rm vir}\approx 10^{14}M_{\odot}$ per
${\rm deg}^2$ per $1\%$ bandwidth for $600\,{\rm MHz}<f_{\rm
obs}<1200\,{\rm MHz}$. There are very few objects with $M_{\rm
vir}>2\times 10^{13}M_{\odot}$ for $f_{\rm obs}=400\,{\rm
MHz}$. Larger objects with $M_{\rm vir}\approx 3\times
10^{14}M_{\odot}$ are much more rare for $f_{\rm obs}=600\,{\rm MHz}$,
but for higher frequencies there would still be many objects above
this limiting mass. Using the fiducial value of $M_{\rm
lim}=10^{14}M_{\odot}$ at $f_{\rm obs}=1000\,{\rm MHz}$, by referring
back to table~\ref{tab:tab2}, we see that it would be possible to
cover $10000\,{\rm deg}^2$ in $\approx 400$ days using setup II. Such
a survey would find $\approx 15000$ objects above this mass limit
since there are $\approx 0.3$ objects per ${\rm deg}^2$ per $1\%$
bandwidth. We note that a similar survey would take only 90 days with
setup III and there would be twice as many objects found since it has
a larger bandwidth. For the same integration time one could cover only
about $1000\,{\rm deg}^2$ at $f_{\rm obs}=800\,{\rm MHz}$ and would
find $\approx 1000$ objects with with setup II. At $f_{\rm
obs}=1200\,{\rm MHz}$ one would be able to cover more area, but there
are less objects objects since the comoving volume is decreasing, and
things are likely to be somewhat more complicated since
$10^{14}M_{\odot}$ objects would be resolved at this frequency. Much
less area could be covered to the same depth at $f_{\rm obs}=600\,{\rm
MHz}$ and as we will discuss it might be that such objects are
difficult to detect against the background since the beamsize
increases with decreasing frequency. Clearly, there is no point in
searching for virialized objects at $f_{\rm obs}=400\,{\rm MHz}$.  

We have also computed the same quantity as a function of $f_{\rm obs}$
in fig.~\ref{fig:dn} for different values of $M_{\rm lim}$. We see
that there are very few virialized objects accessible to observations
with $f_{\rm obs}<600\,{\rm MHz}$ again suggesting that surveys which
are intending to search for such  objects are unlikely to find
anything significant for $z>1.4$. However, we also see that for the
fiducial value of $M_{\rm lim}\approx 10^{14}M_{\odot}$ which we have
been using, there are a wide range of frequencies for which one will
find more than one object per $10\,{\rm deg}^2$ in a $1\%$
bandwidth. Many smaller objects ($M_{\rm vir}\approx 3\times
10^{12}M_{\odot}-10^{13}M_{\odot}$) are accessible at lower
frequencies due to the evolution of structure.  

\subsection{Optimal design and survey strategy}

\begin{figure}
\epsfig{file=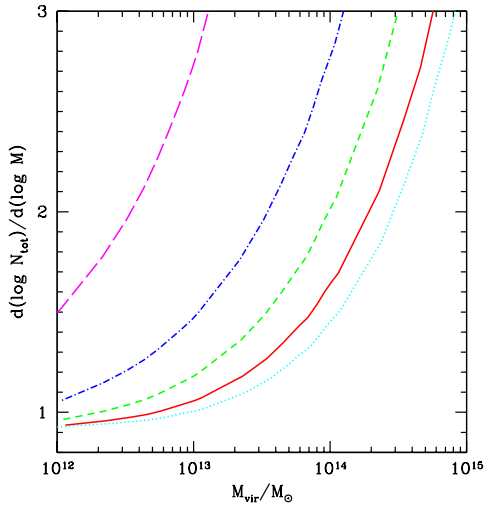,height=7cm}
\caption{The power law $n$ such that $(N/1{\rm deg}^2)\propto M_{\rm
vir}^{-n}$. Under the assumption that $f_{\hi}$ is independent of
$M_{\rm vir}$ the optimal strategy is when $n=5/3$. The lines are
labelled as in fig.~\ref{fig:hi}.} 
\label{fig:nn}
\end{figure}

One can attempt to gain some understanding of the optimal design of an
instrument and survey strategy by substituting into (\ref{ndef}) for
$\Delta\Omega$ from (\ref{rdef}), for $\Delta z$ from (\ref{dz}), and
using $S_{\sigma}\Delta {\rm v}\propto M_{\hi}/d_{L}^2$. One finds that
$N\propto t{\cal M}(M_{\hi}^2/\Delta {\rm v})$ where 
\be 
{\cal M}\propto
{\eta^2A^2\Omega_{\rm inst}\Delta f_{\rm inst}\over T^2_{\rm sys}}\,,
\ee
can be thought of as 
the instrumental figure of merit which clearly has sensible
dependences on the relevant parameters. This effectively quantifies
how many 
objects one would find in a survey ignoring any
potential systematics. Alternatively, the integration time required to
find a fixed number of objects is $\propto {\cal M}^{-1}$. It is clear
that similar arguments can be made for a galaxy redshift survey and
this would, therefore, also apply to the SKA and other similar
instruments. This formula should also yield important information when
applied to  surveys for other kinds of objects.  

The value of ${\cal M}\approx 27$ for the ATA
configuration discussed earlier and its value is presented in
table~\ref{tab:tab1} for the phased arrays.
This figure of merit 
has been  normalized to the theoretical sensitivity of the Parkes
multi-beam (PM) receiver~\citep{Barnes:2001a}, although we note that this
instrument can only observe in a narrow frequency range around
21cm. We have also computed this quantity for the Lovell Telescope
multi-beam receiver and the Green Bank Telescope (single beam); they
are ${\cal M}\approx 0.34$ and 0.21 respectively. Note that we have
used the frequency independent $T_{\rm sys}^{\rm rec}$ to compute this
quantity.  

We see that
the setup I is about a factor of 5 more powerful than the PM, while both the
presented ATA configuration and setup II improve on this substantially. Setup 
III is more than 500 times more powerful than the PM. For a
phased array configuration $\Omega_{\rm inst}\propto n_{\rm b}d^{-2}$
and hence ${\cal M}\propto\eta_{\rm F}=A/d^2$ which represents the filling
factor of the array. It is clear that, at least for unresolved
detections and ignoring the issue of confusion, which we will discuss
below, a totally filled array $(\eta_{\rm F}=1)$
will perform the best in terms of
having the largest number of objects above the detection threshold. This is because an array with $\eta_{\rm F}\ll 1$ has a much lower sensitivity to surface brightness temperature, than one with $\eta_{\rm F}\sim 1$.  Such an array would be likely
to have poor resolution particularly at low frequencies and would,
therefore, make it difficult to make any sub-selection within the
sample to cut down on systematics. It is possible that low resolution arrays could also suffer from confusion related issues (see below).

Assuming a virialized halo $\Delta {\rm v}\propto \sigma\propto M_{\rm
vir}^{1/3}$ ($\sigma$ is the velocity dispersion of the object) and if
$f_{\hi}$ is independent of $M_{\rm vir}$, then the multiplicative
factor $M_{\hi}^2/\Delta {\rm v} \propto M_{\rm vir}^{5/3}$. Since the
integral in (\ref{ndef}) is a decaying function of $M_{\rm vir}$ with
negative power law $n\equiv n(M_{\rm vir})$, the optimal observing
strategy would be to set the noise so that $M_{\rm lim}$ is that for
which $n=5/3$, that is $M_{\rm lim}\approx 2\times 10^{14}M_{\odot} $,
$10^{14}M_{\odot}$, $4\times 10^{13}M_{\odot}$, $1.5\times
10^{13}M_{\odot}$ and $10^{12}M_{\odot}$ for $f_{\rm obs}=1200\,{\rm
MHz}$, $1000\,{\rm MHz}$, $800\,{\rm MHz}$, $600\,{\rm MHz}$ and
$400\,{\rm MHz}$ respectively, this being a simple consequence of the
evolution of structure. We have presented this power law  $n=-d(\log
N)/d(\log M)$ as a function of $M_{\rm vir}$ in fig.~\ref{fig:nn}. We
see that under these assumptions $M_{\rm lim}\approx 10^{14}M_{\odot}$
should be close to the optimal mass limit for $f_{\rm obs}=1000\,{\rm
MHz}$ and  that lower limits, requiring deeper surveys and hence less
angular coverage for a fixed integration time, are needed  to be
optimal at lower frequencies. 
We should  caution that this is heavily predicated on the assumption
that $f_{\hi}$ is independent of $M$ which we have already argued is
unlikely to be the case and this would have to be taken into account
before relying on such a calculation to set the depth of an actual
observational strategy. If $f_{\hi}$ decreases with $M$ then this
would mean one should perform a deeper survey than one would if it
were independent of $M$. Suffice to say, if one has some idea as to
the dependence of $f_{\hi}$ on $M$, the method would remain the same,
but with a different value of $n$. 

\subsection{Confusion noise}

\begin{figure}
\epsfig{file=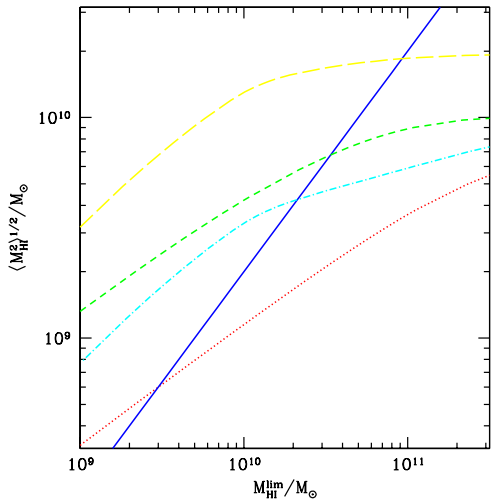,height=7cm}
\caption{An attempt to estimate the confusion noise as function of the
limiting \hi mass of a survey, $M_{\hi}^{\rm lim}$. The solid line is
a representation of the thermal noise: when the other curves are below
it the confusion noise is less than the contribution due to thermal
noise. The dotted line is for $z=0.42$ which corresponds to $f_{\rm
obs}=1000\,{\rm MHz}$ with $f_{\hi}=10^{-3}$ for all $M$. The
equivalent curve for $z=1.37$ ($f_{\rm obs}=600\,{\rm MHz}$) is the
short dashed line. The dashed-dotted line ($z=0.42$) and the long
dashed line ($z=1.37$) are those given by using $f_{\hi}$ from
(\ref{fhi}). The values of $\theta_{\rm FWHM}$ are those used for
setup II.} 
\label{fig:conf}
\end{figure}

Observations using spectral lines are not typically effected by issues
of confusion as can often be the case for continuum sources. However,
in the current situation we are dealing with very large amounts of
$\hinsp$ and very large beams. One has, therefore, to be careful to avoid
making the beam size so large that a typical beam contains an amount
of \hi comparable to the detection threshold. This becomes more and
more important as one makes deeper and deeper surveys, particularly at
high redshift. As we shall see this  is 
very sensitive to the small scale distribution of \hinsp, but in order
to make a simple estimate let us note that the comoving volume
enclosed by a beam of 
$10^{\prime}$ and a velocity width of $800\,{\rm km}\,{\rm sec}^{-1}$ at
$z\approx 1$ is $\sim 1000\,{\rm Mpc}^3$. Since the comoving matter density
is $\approx 4\times 10^{10}M_{\odot}{\rm Mpc}^{-3}$, this means that a typical
beam volume contains about $4\times 10^{13}M_{\odot}$ of dark matter and
using our estimate for $f_{\hi}$, about $4\times 10^{10}M_{\odot}$ of
$\hi$. Clearly this mitigates our earlier assertion that $\eta_{\rm
F}\approx 1$ would be the best situation for unresolved
detections. Making $\theta_{\rm FWHM}$ sufficiently small as to avoid
this issue is clearly another important design criterion. 

An interferometer or phased array would not, in fact, be sensitive to
this smooth mass distribution, but rather to the fluctuations in it,
which are typically smaller. Therefore, the discussion above yields an
over-estimate of the possible effects of confusion. The clusters have
overdensities of $\sim 100$ in the dark matter and we have already
pointed out that, so far as current observations can tell, the value
of $f_{\hi}$ in the field is only marginally larger than that in a
cluster, for example, in A3128. Even if the overdensity in \hi were to
be diluted to only $\sim 20$, one should not have any problem in
picking them out from the background  with a telescope whose beam is
approximately the same size as the object. Velocity structure can only
help in this respect. The problem is that for a given telescope the
resolution degrades very rapidly as redshift increases. Hence, if a
telescope were to be ideally suited to detection of clusters at
$z\approx 0.5$ then by $z\approx 1$ this would lead to a dilution of
the overdensity by factor of $\sim 4$. 

One can make an estimate of the rms fluctuation in the mass in a
volume defined by the beam area and the velocity width $\Delta{\rm
v}=800\,{\rm km}\,{\rm sec}^{-1}$ by computing 
\be
\langle M_{\hi}^2\rangle=\Delta z \Delta\Omega{dV\over
dz\,d\Omega}\int_{M_{\rm lim}}^{\infty} 
[f_{\hi}(M,z)]^2M^2{dn\over dM}dM\,,
\ee
where $\Delta z =(1+z)d{\rm v}=(1+z)f_{\rm obj}/f_{\rm obs}$ is the
velocity width of the object and $\Delta\Omega\propto \theta_{\rm
FWHM}^2$ is the beam area. We see that in order to estimate the
confusion noise $\langle M^2_{\hi}\rangle^{1/2}$ one needs to know
$f_{\hi}(M,z)$ for all $M$ at the particular redshift in question,
that is, we need to extrapolate our argument for $f_{\hi}$ down to the
galactic scale and below, which is beyond the scope of this paper. We
present the results for an experimental configuration similar to setup
II, that is, $z\approx 0.42$ ($\theta_{\rm FWHM}\approx 5^{\prime}$)
and $z\approx 1.37$ ($\theta_{\rm FWHM}\approx 8.4^{\prime}$) using
the fixed value of $f_{\hi}\approx 10^{-3}$ in fig.~\ref{fig:conf},
along with some attempt to interpolate between $f_{\hi}\approx
10^{-2}$ expected for lower mass objects and that which we have argued
applies to larger objects, $f_{\hi}\approx 10^{-3}$. We do this in an
ad hoc way using the function  
\be
f_{\hi}(M,z)={10^{-3}(M/M_{\odot})+10^{12}\over (M/M_{\odot})+10^{14}}\,,
\label{fhi}
\ee
to gain some insight into what the possible effects could be. We
stress that we are not claiming that this expression has any physical
origin, apart from the fact that it models the correct kind of
behaviour. For $M\gg 10^{10}M_{\odot}$ this function yields
$f_{\hi}\approx 10^{-3}$ and for $M\ll10^{10}M_{\odot}$, it yields
$f_{\hi}=10^{-2}$. We note that in such a  model the overall \hi
content of the universe is larger than in one with the fixed value of
$f_{\hi}=10^{-3}$. 

We see that even when using (\ref{fhi}) observations for setup II at
$f_{\rm obs}=1000\,{\rm MHz}$, one would have a confusion noise which
is much lower than the thermal noise mass limit if $M_{\hi}^{\rm
lim}>2\times 10^{10}M_{\odot}$. Using the constant value of
$f_{\hinsp}=10^{-3}$ the restriction is even weaker. Therefore, we can
conclude that the confusion noise would be very much lower than the
thermal noise value for the fiducial value of $M_{\hi}^{\rm
lim}=10^{11}M_{\odot}$ used throughout this paper. As one might expect
things are more restrictive at $f_{\rm obs}=600\,{\rm MHz}$. For the
fixed value of $f_{\hi}$ one would be restricted to $M_{\hi}^{\rm
lim}>3\times 10^{10}M_{\odot}$ and if (\ref{fhi}) were to be true the
confusion noise would be comparable to the thermal noise at
$M_{\hi}^{\rm lim}=10^{11}M_{\odot}$. One would be more likely to be
effected  by confusion noise for setup I due its large beam and less
for setup III and the ATA since they have smaller beams. 

\section{Detecting proto-clusters}

\subsection{Number of dark matter halos at turnaround}

The majority of our discussion to date has focused on the possibility
of detecting virialized objects. However, we have noted that such
objects are likely to have had their \hi content depleted relative to
the field by at least $20-30\%$ during the process of
virialization. It should also be possible to detect objects which are
just beginning collapse, at turnaround, when they are just decoupling
from the Hubble flow. This could be more efficient at high redshift
where there are likely to be very few virialized objects with $M_{\rm
vir}>10^{14}M_{\odot}$ and virialized objects could become confused if
the beam is too large. At turnaround $\Delta_{\rm c}\approx 5$ and by
virtue of the fact that virialization has not yet taken place, it
should be possible to use, with some confidence, the value of
$f_{\hi}$ for the field at that time. Assuming a linear rate of star
formation between $z=0$ and $z=2.4$, one can deduce that 
\be 
f_{\hi}^{\rm field}(z)\approx (1.2+2.8z)\times 10^{-3}\,,
\ee 
for the field.

Moreover it is possible that the velocity dispersion of such objects
is much less than the fiducial value of $\Delta{\rm v}=800\,{\rm
km}\,{\rm sec}^{-1}$ used in the earlier parts of the paper for
virialized objects. If this is so, the signal to noise for a fixed
integration time on a source would increase. This is because  the
signal is $\propto (\Delta{\rm v})^{-1}$ for a fixed \hi mass and the
noise is $\propto (\Delta {\rm v})^{-1/2}$. The typical value of
$\Delta{\rm v}$ for such a object is unknown. Formally, the point of
turnaround is defined to be that when the average velocity is zero,
but the velocity dispersion need not be so. It will be governed by
that of the virialized objects within the region which are in the
process of merging to build up the cluster. In the subsequent
discussion we will allow for $\Delta{\rm v}$ to be a modelling
parameter, but it is  worth discussing various possible values that it
might take. If the object at turnaround just comprises of two large
objects which are collapsing together then the value of $\Delta{\rm
v}$ will be close to the fiducial value of $800\,{\rm km}\,{\rm
sec}^{-1}$. It is possible that such an object contains
$10^{14}M_{\odot}$ of dark matter and is comprised of $\sim 100$
smaller galactic-scale objects in the process of infall with masses of
$\sim 10^{12}M_{\odot}$. In this case $\Delta{\rm v}\sim 100-200\,{\rm
km}\,{\rm sec}^{-1}$. Or it could be that there are many more smaller
objects with, say, $\Delta{\rm v}\sim 20-50\,{\rm km}\,{\rm
sec}^{-1}$. These possibilities are worth remembering in the
subsequent discussion. 

We should note that the collapsing objects at  $z\approx 1-2$ are the
proto-clusters which would have virialized by $z\approx 0.5-1$. If
$R_{\rm ta}$, is the radius of the object at turnaround then the
radius of that object once it has virialized is given by $R_{\rm
vir}=R_{\rm ta}/2$ assuming a completely matter-dominated
universe~\citep{sph1a,ram}. Corrections can be made to include a
cosmological constant~\citep{sph2}  and also dark
energy~\citep{sph3a,sph3b}. Similar relations can be derived relating
the time at turnaround $t_{\rm ta}$ and that at virialization $t_{\rm
vir}=2t_{\rm ta}$, and the corresponding redshifts, $1+z_{\rm
ta}=2^{2/3}(1+z_{\rm vir})$. \citet{LL} give  a detailed exposition of
this model.  

The formula (\ref{dndM}) applies to objects with an overdensity,
$\Delta_{\rm c}=200$. If one assumes an NFW profile function for the
objects with $c=5$, one can show that  $M_{200}\approx M_{5}/2$; this
relation is approximately true for a wide range of concentration
parameters. Taking the fiducial value of $M_{\hi}^{\rm
lim}=10^{11}M_{\odot}$ and $f_{\hi}\approx 5\times 10^{-3}$ for
$f_{\rm obs}=600\,{\rm MHz}$, we see that $M_{5}^{\rm lim}\approx
2\times 10^{13}M_{\odot}$ and the corresponding value of
$M_{200}\approx 10^{13}M_{\odot}$ which can  be used in conjunction
with figs.~\ref{fig:hi}, \ref{fig:dn} and \ref{fig:nn} to deduce the
number of collapsing objects that would be found in a given survey, if
one ignores the smaller correction from $M_{200}$ to $M_{\rm vir}$. We
see that there are $\approx 20$ objects of this size per ${\rm deg}^2$
per $1\%$ bandwidth. For $f_{\rm obs}=400\,{\rm MHz}$ the value of
$f_{\hi}$ is higher and hence there are about the same number of
objects; the increase in $f_{\hi}$ offsetting the smaller number of
objects for a given $M_{\rm vir}$. With this large number of objects
and the large beams likely at this observing frequency, one might
think that one would be close to confusion limited ($\sim 1$ object
per beam area), but the extra information provided by the velocity
information should help avoid this possibility.  

Given the larger value of $f_{\hi}$ likely for objects which have not
virialized, it might be sensible to also consider the possibility of
$M_{\hi}^{\rm lim}=10^{12}M_{\odot}$, for which the corresponding
values of $M_5^{\rm lim}$ and $M_{200}^{\rm lim}$ are 10 times
larger. We see that there are still numerous objects ($\sim 0.1$ per
${\rm deg}^2$ per $1\%$ bandwidth) of this size for $f_{\rm
obs}=600\,{\rm MHz}$, although there are markedly less for lower
values of $f_{\rm obs}$ even taking into account the larger value of
$f_{\hi}$. Achieving this larger limiting mass at $600\,{\rm MHz}$
should be possible in a fraction of the time required for
$10^{11}M_{\odot}$. 

\begin{figure}
\epsfig{file=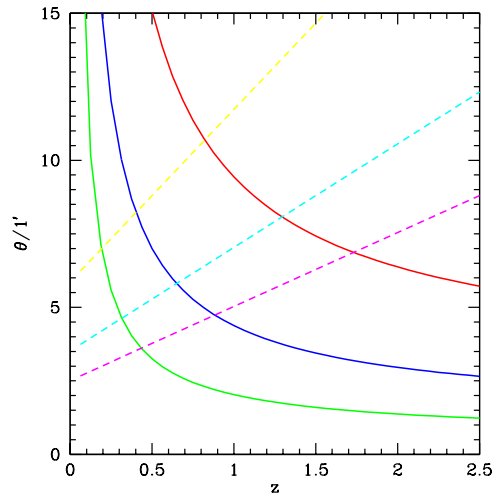,height=7cm}
\caption{The angular diameter size (twice the radius) of  objects
which are at turnaround  with $M_{5}=10^{14}M_{\odot}$,
$10^{13}M_{\odot}$ and $10^{12}M_{\odot}$ (solid lines, top to bottom
respectively). Included also is the estimated beam size for the three
different phased array configurations (dashed lines, I top, II middle,
III bottom). Note that the angular size of the collapsing objects are
much larger than the objects of the same virial mass at the same
redshift.} 
\label{fig:thetap}
\end{figure}

We should note that at any given redshift the objects at turnaround
will be larger than those which are virialized, but have the same
overall mass. Fig.~\ref{fig:thetap} shows the angular diameter size of
object with a variety of masses at turnaround. We see that these
objects are larger than virialized objects of the same mass at the
same redshift. At the high redshifts we are considering here, the
objects still have a angular diameter  less than $5^{\prime}$ and
hence they are likely to unresolved by the instruments under
discussion. At lower redshifts such objects would be resolved and it
should be more efficient to detect virialized objects. 

\subsection{Detection related issues}

We can adapt the earlier sensitivity calculation to an arbitrary value
of $\Delta{\rm v}$ and $M_{\hi}^{\rm lim}$. In particular one can
compute the ${\cal R}_{\Delta{\rm v}}$, $S_{\Delta{\rm v}}$ and
$t_{\Delta{\rm v}}$ from ${\cal R}_{800}$, $S_{800}$ and $t_{800}$. We
see that the survey rate is given by  
\be 
{\cal R}_{\Delta{\rm v}}={\cal R}_{800}\left({800\,{\rm km}\,{\rm
sec}^{-1}\over \Delta{\rm v}}\right)^{1/2}\,. 
\ee
For an \hi mass of $M_{\hi}^{\rm lim}$ the required flux density is 
\be 
S_{{\Delta{\rm v}}}=S_{800}\left({800\,{\rm km}\,{\rm sec}^{-1}\over
\Delta{\rm v}}\right)\left({M_{\hi}^{\rm lim}\over
10^{11}M_{\odot}}\right)\,, 
\ee
and the actual integration time required to achieve a $5\sigma$
detection of such an object on an area $100\,{\rm deg}^2$ is 
\be
t_{\Delta{\rm v}}=t_{800}\left({\Delta{\rm v}\over 800\,{\rm km}\,{\rm
sec}^{-1}}\right)\left({10^{11}M_{\odot}\over M_{\hi}^{\rm
lim}}\right)^2\,. 
\ee

If we assume that $\Delta{\rm v} = 100\,{\rm km}\,{\rm sec}^{-1}$ then
we see that it would be possible to cover close to $260\,{\rm deg}^2$
in a day of integration $M^{\rm lim}_{\hi}=10^{12}M_{\odot}$ using
setup II at $f_{\rm obs}=600\,{\rm MHz}$. Since there are $\approx
0.15$ objects per ${\rm deg}^2$ per $1\%$ bandwidth above the
corresponding mass limit then one would hope to find around 200
collapsing objects. It would take around 10 times as long to do a
similar survey at $f_{\rm obs}=400\,{\rm MHz}$ and one would find
approximately 10 times fewer objects taking into account the larger
value of $f_{\hi}$ and the very much reduced number of objects with a
given mass. Nonetheless, $\sim 20$ objects in 10 days of integration
time is definitely worthwhile. 

Neither of the above survey parameters are optimal. If we assume that
$\Delta{\rm v}$ is either weakly dependent on $M_{5}$ or not at all,
then the optimal mass limit would correspond to the value of $M_{200}$
for which $n=2$ in contrast to the virialized case. Assuming that
$M_{200}\approx M_{\rm vir}$, we see from fig.~\ref{fig:nn} that the
optimal values of $M_{200}$ are $3 \times 10^{13}M_{\odot}$ and
$2\times 10^{12}M_{\odot}$ for $f_{\rm obs}=600\,{\rm MHz}$ and
$400\,{\rm MHz}$ respectively. The corresponding \hi mass limits are
$M_{\hi}^{\rm lim}=3\times 10^{11}M_{\odot}$ and $3\times
10^{10}M_{\odot}$. For $f_{\rm obs}=600\,{\rm MHz}$, it would require
4.3 days of integration time to achieve this optimum depth on
$100\,{\rm deg}^2$. There are $\approx 2.3$ objects per ${\rm deg}^2$
per $1\%$ bandwidth and hence such a survey would find $\approx 1150$
objects (which is more than the $\approx 860$ that would  be found by
mapping  $1110\,{\rm deg}^2$ in 4.3 days with a limit of
$10^{12}\,M_{\odot}$ as suggested above). 

\section{Conclusions}

To summarise, we have shown that instruments likely to be built within the
next few years have a realistic chance of detecting large objects, both
virialized and collapsing, using $\hi$ emission as their tracer
opening a new window on the universe. If a detection threshold of
$M_{\hi}^{\rm lim}\approx 10^{11}M_{\odot}$ can be achieved at around
$z\approx 0.4$ then it should be possible to find a surprizingly large
number of virialised objects. Similarly, it should be possible to
detect many objects at turnaround with $z>1$. We have also made comments
as to the optimal design of an instrument and the survey strategy for
these applications. Clearly, more sophisticated simulations of the
large-scale distribution of \hi  are required, but  we believe that
the basic picture we have put forward is likely to remain.

It is clear that the detection of the large number of objects, both
virialized and collapsing, predicted in this paper could have a
significant impact on our understanding of the universe. In the regime
where one can optimally detect virialized clusters ($z<1$) it should
be possible, using the extra velocity information, to accurately
compute the dark matter mass of each of the objects which are detected
and establish the selection function. Since the number of virialized
objects is sensitive to cosmological parameters, accurate estimates of
$\Omega_{\rm m}$ and $\sigma_8$ should be possible. Moreover, the
properties of the dark energy may also be accessible to such an
analysis. The nature of the collapsing structures for $(z>1)$ is also
of significant interest. We have used all the available information to
make estimates of the number of objects which would be found. However,
we have also noted that these are somewhat uncertain, particularly the
velocity structure. Clearly the detection of a large number of objects
will have a significant impact on our understanding of the
distribution of \hi at high redshifts and the on-going  process galaxy
formation. 
  
\section*{Acknowledgements}
We would like to thank Martin Rees, Martin Haehnelt, Ian Browne and
Peter Wilkinson for extremely helpful comments and suggestions. RAB is
funded by PPARC and JW in part by Kings College. 

\bibliographystyle{mn2e}

\label{lastpage}
\end{document}